\documentclass[12pt,preprint]{article}
\usepackage{amsmath}
\usepackage{graphics}
\usepackage{epsfig}
\renewcommand{\vec}[1]{{\bf #1}}
\newcommand{\eqb}{\begin{equation}}
\newcommand{\eqe}{\end{equation}}
\newcommand{\dmb}{\begin{displaymath}}
\newcommand{\dme}{\end{displaymath}}
\newcommand{\pd}{\partial}

\newcommand{\eab}{\begin{eqnarray}}
\newcommand{\eae}{\end{eqnarray}}

\newcommand{\be}{\begin{equation}}
\newcommand{\ee}{\end{equation}}

\newcommand{\La}{\Lambda}
\begin{document}
\begin{titlepage}
\begin{flushright}
\end{flushright}
\vspace{0.6cm}

\begin{center}
\Large{Yang-Mills thermodynamics: The preconfining phase}

\vspace{1.5cm}

\large{Ralf Hofmann}

\end{center}
\vspace{1.5cm}

\begin{center}
{\em Institut f\"ur Theoretische Physik\\ 
Universit\"at Frankfurt\\ 
Johann Wolfgang Goethe - Universit\"at\\ 
Robert-Mayer-Str. 10\\ 
60054 Frankfurt, Germany}
\end{center}
\vspace{1.5cm}

\begin{abstract}
We summarize recent nonperturbative results obtained for the 
thermodynamics of an SU(2) and an SU(3)
Yang-Mills theory being in its 
preconfining (magnetic) phase. We focus 
on an explanation of the involved concepts and 
derivations, and we avoid technical details. 

\end{abstract} 

\end{titlepage}

\noindent{\sl Introduction.} This is the second one 
in a series of three papers summarizing the 
thermodynamics of an SU(2) and an SU(3) Yang-Mills theory as it is analysed 
nonperturbatively in \cite{Hofmann2005}. The goal is to give 
a nontechnical presentation of the concepts involved in 
quantitatively describing the preconfining 
or magnetic phase. In the phase diagram of either theory 
this phase is sandwiched inbetween the deconfining phase (bosonic statistics) and 
the absolutely confining (fermionic statistics) phase. 
While fundamental, heavy, and fermionic test charges 
are confined by condensed magnetic monopoles dual gauge 
modes are massive but propagate.       
  
To derive the phenomenon of quark confinement from a 
microscopic Lagrangian based on an SU(3) gauge 
symmetry, that is, from Quantum Chromodynamics (QCD), 
is a major challenge 
to human thinking. This is due to the 
theory being strongly interacting at large distances which 
disqualifies a perturbative approach 
to this problem. The difficulties are overwhelming even in 
the somewhat simplified situation, where quarks are considered 
heavy and nondynamical test charges that are immersed into pure gluodynamics. 
 
Two main proposals for the confining mechanism are 
discussed in the literature: The dual Meissner 
effect, which takes place within a condensate 
of massless magnetic monopoles \cite{dualsuperconductor}, 
and the condensation of infinitely mobile magnetic center-vortex loops 
\cite{centervortexcondensate} implying a dielectricity of the ground state 
which strongly increases with distance. In its electrically dual form 
the former mechanism is observed in Nature if a superconducting material is 
subjected to an external, static magnetic field. 
Within the Cooper pair condensate 
the field lines of the latter are forced 
into thin flux tubes \cite{GinzburgLandauAbrikosov19501957}: an immediate consequence of 
infinitely mobile, that is, stiffly correlated electric charges. 
In the hypothetical situation, where the magnetic field is sourced by a pair of a 
heavy magnetic monopole and antimonopole at distance $R$ the squeezing-in 
of flux lines leads to a linear 
potential at large $R$ and thus to monopole 
confinement. A similar situation would hold if the 
condensate of Cooper pairs would be replaced 
by infinitely mobil magnetic dipoles. The magnetic dual of this 
phenomenon is that a condensate of electric dipoles 
confines a heavy electric charge and 
its anticharge. But an electric dipole originates 
from a circuit of magnetic flux. In an SU(N) Yang-Mills theory the 
latter is naturally provided by a 
magnetic center-vortex loop. The confinement 
mechanism involving a dual superconductor 
and a center-vortex condensate are mutually exclusive. 
One of the results in \cite{Hofmann2005} is, however, that either of 
the two mechanisms belongs 
to one of the two separate phases with 
test-charge confinement in SU(2) and SU(3) 
gluodynamics.  
    
The objective of the present paper is the discussion of a 
phase of SU(2) and SU(3) gluodynamics whose 
ground state is a dual superconductor (magnetic phase). We focus on 
the SU(2) case and mention generalizations to the SU(3) case and 
the associated results in passing only. 
In discussing the properties of the ground state 
and those of its (noninteracting) quasiparticle 
excitations we pursue the following program: 

First, we remind the reader why pairs of nonrelativistic and screened Bogomol'nyi-Prasad-Sommerfield (BPS) \cite{PS1974} magnetic monopoles and antimonopoles arise as spatially 
isolated defects in the deconfining (electric) phase. 

Second, a continuous parameter, namely, the magnetic flux 
originating from a zero-momentum pair of a monopole and 
its antimonopole, both located at spatial infinity, is computed in 
the limit of total screening. The limit of total magnetic charge screening 
is dynamically reached at the electric-magnetic phase boundary \cite{Hofmann2005,Hofmann2005_1}. 
Recall, that total screening means masslessness for each 
individual monopole or antimonopole and that the projection onto 
zero three-momentum corresponds to a spatial coarse-graining. 
The obtained magnetic flux, being an angular 
parameter, is identified with the euclidean time $\tau$. Since only charge-modulus one 
monopoles and antimonopoles contribute to the flux the winding number, associated with 
the phase of a complex scalar field $\varphi$, is $\pm$ unity. In the 
(at this stage hypothetical but in a later, intermediate step shown to be selfconsistent) 
absence of interactions between the monopoles and antimonopoles in the condensate the 
field $\varphi$ is energy- and pressure-free. That is, its euclidean 
time dependence is BPS saturated. Assuming the existence of an externally 
provided mass scale $\Lambda_M$, this determines $\varphi$'s BPS equation. The latter, in turn, 
uniquely fixes $\varphi$'s potential $V_M$. Similar to the case of the adjoint scalar 
field $\phi$ in the electric phase the statistical and quantum mechanical 
inertness of the complex field $\varphi$ is established by comparing $V_M$'s 
curvature with the squares of temperature $T$ and 
the maximal resolution $|\varphi|$. Notice that $|\varphi|$ 
is the maximal spatial resolution after a coarse-graining up to a 
length scale $|\varphi|^{-1}$ has been performed.    

Third, having derived the coarse-grained monopole physics 
in the absence of interactions the full effective theory is obtained by a 
minimal coupling of coarse-grained dual plane waves to the inert 
monopole sector. (The electric-magnetic transition is survived 
only by gauge fields transforming under $U(1)_D$ (SU(2)) and 
$U(1)^2_D$ (SU(3)) \cite{Hofmann2005}). A pure-gauge solution 
$a^{D,bg}_\mu$ to the dual gauge-field equations of motion emerges 
in the effective theory. This is the coarse-grained manifestation of 
monopole and antimonopole interactions 
mediated by plane-wave quantum fluctuations in the dual gauge fields. 
(The averaged-over quantum fluctuations are of 
off-shellnes larger than $|\varphi|^2$.) 
By virtue of the pure-gauge configuration $a_D^{bg}$ the vanishing pressure 
and the vanishing energy density of a 
(hypothetical) condensate of noninteracting monopoles and 
antimonopoles are shifted to $\rho^{gs}=-P^{gs}=\pi\,\La_M^3 T$ 
(SU(2)). (For SU(3) two monopole condensates $\varphi_1$, $\varphi_1$ and 
two pure-gauge configurations $a^{D,bg}_{\mu,1}=a^{D,bg}_{\mu,2}$ exist, 
and one has $\rho^{gs}=-P^{gs}=2\pi\,\La_M^3 T$.) The negative ground-state presssure 
can, on a microscopic level and at finite magnetic coupling $g$, be 
understood in terms of magnetic flux loops which collapse as soon as the are 
created. The collapse takes place under the influence of negative 
pressure occuring away from the vortex cores \cite{Hofmann2005}. The coarse-grained  
manifestation of dual gauge modes scattering off the magnetic monopoles 
or antimonopoles along a zig-zag like path within the condensate 
is the abelian Higgs mechanism. The latter gives rise to quasiparticle masses. 

Fourth, the required invariance of the Legendre 
transformations for thermodynamical quantities under 
the applied coarse-graining yields a first-order differential 
equation which governs the evolution of the magnetic coupling 
$g$ with temperature. This evolution allows to 
analyse the electric-magnetic phase transition and 
the transition to the absolutely confining (center) 
phase where center-vortex loops emerge as fermionic particles 
from the decaying ground state. Moreover, 
it determines the evolution of thermodynamical quantities. This evolution is 
exact on the one-loop level.\vspace{-0.3cm}\\ 

\noindent{\sl Screened BPS magnetic monopoles.} To understand the origin 
of isolated and screened magnetic 
monopoles and antimonopoles in the 
deconfining phase \cite{KorthalsAltes} 
we recall some properties of certain, 
topologically nontrivial, (anti)selfdual solutions 
to the euclidean Yang-Mills equations 
at finite temperature: calorons and anticalorons 
of nontrivial holonomy, topological 
charge modulus $|Q|=1$, and no net magnetic charge \cite{nthcalorons}. 
As it turns out \cite{Hofmann2005,HerbstHofmann2004} only these configurations are relevant for a coarse-grained re-formulation of the fundamental theory at high temperatures. 
(The term holonomy refers 
to the behavior of the Polyakov loop when evaluated on a 
gauge-field configuration at spatial infinity. A configuration is associated with a 
trivial holonomy if its Polyakov loop is in the center of the group ($Z_2$ for SU(2) and 
$Z_3$ for SU(3)) and associated with a nontrivial holonomy otherwise.) 
On a microscopic level nontrivial holonomy is 
excited out of trivial holonomy by gluon exchanges 
between calorons and anticalorons. On the classical level 
the (anti)selfdual nontrivial-holonomy solution possesses a pair of a 
BPS monopole and its antimonopole, which do not interact, 
as constituents. Switching on one-loop 
quantum fluctuations, a situation investigated for an isolated (anti)caloron in 
\cite{Diakonov2004}, the monopole and its antimonopole either 
attract (small holonomy) or repulse (large holonomy) 
each other. The former process is much more likely than the latter and 
leads to a {\sl negative} ground-state pressure after 
spatial coarse-graining: a monopole and its antimonopole attract each other 
so long until the annihilate and subsequently get re-created elsewhere. 
The repulsion between a monopole and 
its antimonopole, which both originate from a quantum blurred 
large-holonomy caloron, fades with an increasing number of 
small-holonomy caloron fluctuations taking place inbetween 
these particles. This facilitates the life of screened monopoles and antimonopoles 
in isolation. The screening of the 
magnetic charge $g=\frac{4\pi}{e}$ by small-holonomy 
calorons is, on average, described by the gauge 
coupling $e$ in the coarse-grained theory. A fact, important when considering the 
limit $e\to\infty$ below, is that the sum 
%**********
\eqb
\label{massesma}
M_{m+a}\equiv =\frac{8\pi^2}{e\beta}\,.
\eqe
%**********
of the masses of a screened monopole and its screened 
antimonopole, both originating from a 
dissociating large-holonomy caloron, is independent 
of the holonomy \cite{nthcalorons}. Notice that $M_{m+a}\to 0$ for $e\to\infty$ and that 
this limit dynamically takes place at the electric-magnetic transition, where $e\sim -\log(\lambda_E-\lambda_{c,E})$ 
($\lambda_{c,E}\equiv\frac{2\pi T}{\Lambda_E}$), 
through a total screening of the magnetic 
charge of an isolated monopole and its antimonopole 
by intermediate small-holonomy caloron fluctuations. 
\vspace{0.3cm}\\ 
\noindent{\sl Monopole condensate(s) as spatial average(s).} 
Just as performed in the deconfining phase, we need to derive 
the $\tau$-dependence of the phase of a now complex, spatially homogeneous 
field $\varphi$ (SU(2)) or fields $\varphi_1$, $\varphi_2$ (SU(3)) 
describing the monopole condensate(s) 
after spatial coarse-graining down to a resolution $|\varphi|$ (SU(2)) 
or $|\varphi_1|$, $|\varphi_2|$ (SU(3)). Again, we spatially average over 
large (topological) quantum fluctuations provided by massless, noninteracting, 
and stationary monopoles and antimonopoles and, subsequently, 
over quantum fluctuations carried by plane-waves. 

The average needs to 
be performed in a physical (unitary) gauge where the magnetic 
flux emanating of each isolated monopole 
or antimonopole is compensated for by an 
associated Dirac string. A continuous dimensionless parameter, eventually to 
be identified with $\frac{\tau}{\beta}$, arises when considering the magnetic 
flux in the limit $e\to\infty$ which belongs to a pair of 
noninteracting, stationary (with respect to the heat bath) 
monopole and antimonopole situated outside of an $S^2_\infty$ with 
infinite radius $R=\infty$. The latter acts as a heat 
bath to the pair. Notice that a monopole-antimonopole pair situated inside 
the $S^2_\infty$ does not contribute to the flux. 
(To readers having trouble distinguishing inside 
from outside for an $S^2_\infty$ we propose to consider $R<\infty$ first and 
then take the limit $R\to\infty$.)
 
More specifically, 
we are interested in the average flux through $S^2_\infty$ as a 
function of the angle $\delta$ between the monopole's and 
the antimonmopole's Dirac string. If there were no coupling 
to the heat bath of the monopole-antimonopole pair outside of 
$S^2_\infty$ the mean flux (average over the absolute orientation of the Dirac strings) 
would read \cite{Hofmann2005}
%********
\eqb
\label{avflux}
\bar{F}_{\pm}=\pm\frac{\delta}{2\pi}\frac{4\pi}{e}=
\pm\frac{2\delta}{e}\,,\ \ \ \ \ \ (0\le\delta\le\pi)\,. 
\eqe
%******** 
In the limit $e\to\infty$ we have $\bar{F}_{\pm}\to 0$ and 
no continuous parameter determining the $\tau$ dependence of 
$\varphi's$ phase (SU(2)) would arise. After the coupling to the heat bath is switched on 
and after performing a spatial average the mean occupation number for a massless 
monopole-antimonopole pair (with vanishing spatial momenta of its constituents) 
diverges in a such a way that the mean magnetic flux 
$\bar{F}_{\pm,\tiny\mbox{th}}(\delta)$ is finite. (The total spatial 
momentum of the monopole and antimonopole system vanishes such that each 
individual momentum is zero. Notice that in the absence of a dynamically 
generated scale $\Lambda_M$ the volume $V$, over 
which the spatial average is performed, is undetermined: The only available mass scale $T$, 
which could determine $V$, cancels in the Bose-distribution for $\vec{p}\to 0$ 
and $e\to\infty$ since $M_{m+a}$'s explicit $T$ 
dependence is linear, see Eq.\,(\ref{massesma}).) 

Let us show this in a more explicit way. The thermal average to 
be performed is \cite{Hofmann2005}
%*****
\eab
\label{avfluxsys}
\bar{F}_{\pm,\tiny\mbox{th}}(\delta)&=&
\,4\pi\,\int d^3p\,\delta^{(3)}(\vec{p})\, n_B(\beta |E(\vec{p})|)\,\bar{F}_{\pm}\nonumber\\ 
&=&\pm\frac{8\pi\,\delta}{e}\int d^3p\,
\frac{\delta^{(3)}(\vec{p})}{\exp\left[\beta\sqrt{M_{m+a}^2+\vec{p}^2}\right]-1}\,.
\eae
%*********
After setting $\vec{p}=0$ in 
$\left(\exp\left[\beta\sqrt{M_{a+b}^2+\vec{p}^2}\right]-1\right)$ and by
appealing to Eq.\,(\ref{massesma}), the expansion of this term reads
%*********
\eqb
\label{expexpon}
\lim_{\vec{p}\to 0}\left(\exp\left[\beta\sqrt{M_{m+a}^2+\vec{p}^2}\right]-1\right)=
\frac{8\pi^2}{e}\left(1+\frac{1}{2}\frac{8\pi^2}{e}+
\frac{1}{6}\left(\frac{8\pi^2}{e}\right)^2+\cdots\right)\,.
\eqe
%********
The limit $e\to\infty$ can now safely be performed in Eq.\,(\ref{avfluxsys}), 
and we have
%**********
\eqb
\label{avfluxsys,einf}
\lim_{e\to\infty} \bar{F}_{\pm,\tiny\mbox{th}}(\delta)=
\pm\frac{\delta}{\pi}\,,\ \ \ \ \ \ (0\le\delta\le\pi)\,.
\eqe
%********** 
This is finite and depends on the 
angular variable $\delta$ continuously. Now $\frac{\delta}{\pi}$ is a (normalized) angular 
variable just like $\frac{\tau}{\beta}$ is. Thus we may set $\frac{\delta}{\pi}=\frac{\tau}{\beta}$. 
Since  $\varphi$ (SU(2)) is spatially 
homogeneous (spatial average, $\vec{p}\to 0$!) 
its phase $\hat\varphi\equiv\frac{\varphi}{|\varphi|}$ 
depends on $\frac{\tau}{\beta}$ only and this in a periodic way. Moreover, 
since the physical flux situation for the thermalized monopole-antimonopole pair 
does not repeat itself for $0\le \frac{\delta}{\pi}\le 1$ we conclude that 
this period is $\pm$ unity: 
%*********
\eqb
\label{varphiphase}
\hat\varphi(\tau)=\exp\left[\pm 2\pi\frac{\tau}{\beta}\right]\,. 
\eqe
%*********
To derive $\varphi$'s modulus, which together with $T$ determines the length scale 
$|\varphi|^{-1}$ over which the spatial average is performed, 
we assume the existence of an (at this stage) externally 
provided mass scale $\Lambda_M$. Since the weight for integrating 
out massless and noninteracting monopole-antimonopole systems in the partition 
function is $T$ independent and since the cutoff in length for the spatial 
average defining $|\varphi|$ is $|\varphi|^{-1}$ an explicit $T$ 
dependence ought not arise in any quantity being 
derived from such a coarse-graining. That is, in the effective action density any $T$ 
dependence (still assuming the absence of interactions between massless monopoles and 
antimonopoles when performing the coarse-graining) must appear through 
$\varphi$ only. Moreover, since integrating massless and momentum-free monopoles 
and antimonopoles into the field $\varphi$ means that this field is energy- 
and pressure-free $\varphi$'s $\tau$ dependence (residing in its phase) 
must be BPS saturated. On the right-hand 
side of $\varphi$'s or 
$\bar{\varphi}$'s BPS equation the requirement of analyticity (because away from a phase 
transition the monopole condensate should exhibit a smooth $T$ dependence) 
and linearity (because the $\tau$ dependence of $\varphi$'s 
phase, see Eq.\,(\ref{varphiphase}), needs to honoured) 
in $\varphi$ or $\bar{\varphi}$ yields 
the following first-order equation of motion
%*******
\eqb
\label{BPSvarphi}
\pd_\tau \varphi=\pm i\frac{\La_M^3\,\varphi}{|\varphi|^2}=\pm i\frac{\La_M^3}{\bar{\varphi}}\,.
\eqe
%*******
Notice that Eq.\,(\ref{BPSvarphi}) is not invariant 
under Euclidean boosts: A manifestation of the existence of a 
singled-out rest frame -- the spatially isotropic and spacetime homogeneous 
heat bath. Substituting $\varphi=|\varphi|\hat\varphi$ into Eq.\,(\ref{BPSvarphi}) 
and appealing to Eq.\,(\ref{varphiphase}), 
we derive $|\varphi|=\sqrt{\frac{\La_M^3\beta}{2\pi}}$. 
Notice that the `square' of the right-hand side in Eq.\,(\ref{BPSvarphi}) uniquely 
defines $\varphi$'s potential $V_M$. (In contrast to a second-order equation of 
motion, following from an action by means of the variational 
principle, Eq.\,(\ref{BPSvarphi}) does {\sl not} 
allow for a shift $V_M\to V_M+\mbox{const}$.) For the case SU(3) a BPS equation 
(\ref{BPSvarphi}) arises for each of the two independent monopole 
condensates $\varphi_1$ and $\varphi_2$. 

Finally, one shows by comparing the curvature of their 
potentials with the square of temperature and the squares of 
their moduli that the field $\varphi$ (SU(2)) and the fields $\varphi_1$, 
$\varphi_2$ (SU(3)) do neither fluctuate on-shell nor off-shell, 
respectively \cite{Hofmann2005}: Spatial coarse-graining 
over nonfluctuating, classical configurations generates 
nonfluctuating macroscopic fields.\vspace{-0.3cm}\\ 

\noindent{\sl Effective theory: Thermal ground state and 
dual quasiparticles.}
To obtain the full effective theory the spatially coarse-grained and topologically 
trivial (dual) gauge fields $a^D_\mu$ (SU(2)) and $a^D_{\mu,1}$, $a^D_{\mu,2}$ (SU(3)) are 
minimally coupled (with a universal {\sl effective} magnetic coupling $g$) 
to the inert fields $\varphi$ (SU(2)) and 
$\varphi_1$, $\varphi_2$ (SU(3)). Since the effective theory is abelian 
with (spontaneously broken) gauge group U(1)$_D$ (SU(2)) and U(1)$^2_D$ (SU(3)) 
and since the monopole fields do not fluctuate thermodynamical quantities are exact on the 
one-loop level. Before we discuss the 
spectrum of quasiparticles running in the loop 
we need to derive the full ground-state 
dynamics in the effective theory. The classical equations of motion for 
the dual gauge field $a^D_\mu$ are
%*********
\eqb
\label{eomdualG2}
\pd_\mu G^D_{\mu\nu}=ig\left[\overline{{\cal D}_{\nu}\varphi}\varphi-\bar{\varphi}
{\cal D}_{\nu}\varphi\right]
\eqe
%*********
where $G^D_{\mu\nu}=\pd_\mu a^D_\nu-\pd_\nu a^D_\mu$ and 
${\cal D}_{\mu}\equiv \pd_\mu+ig\,a^D_\mu$. (For SU(3) the 
right-hand sides for the two equations for the dual gauge fields $a^D_{\mu,1}$, $a^D_{\mu,2}$ can be 
obtained by the substitutions $\varphi\to\varphi_1$ or 
$\varphi\to\varphi_2$ in Eq.\,\ref{eomdualG2}.) The pure-gauge solution 
to Eq.\,(\ref{eomdualG2}) with ${\cal D}_\mu\varphi\equiv0$ is given as
%*******
\eqb
\label{pgsolM2}
a^{D,bg}_{\mu}=\pm\delta_{\mu 4}\frac{2\pi}{g\beta}\,.
\eqe
%******** 
In analogy to the deconfining phase, the coarse-grained 
manifestation $a^{D,bg}_{\mu}$ of monopole-antimono\-pole in\-teractions, 
mediated by dual, off-shell plane-wave modes on 
the microscopic level, shifts the energy density $\rho^{gs}$ and the 
pressure $P^{gs}$ of the ground state from zero to finite values: 
$\rho^{gs}=-P^{gs}=\pi\,\La_M^3\,T$ (SU(2)) and 
$\rho^{gs}=-P^{gs}=2\pi\,\La_M^3\,T$ (SU(3)). In contrast to the 
deconfining phase, where the negativity of $P^{gs}$ arises 
from monopole-antimonopole attraction, the negative ground-state pressure 
in the preconfining phase originates from collapsing and re-created 
center-vortex loops \cite{Hofmann2005}. (There are 
two species of such loops for SU(3) and one species for SU(2)). The core of 
a given center-vortex loop can be pictured as a stream of the associated monopole 
species flowing oppositely directed to the stream of their 
antimonopoles \cite{Olejnik1997}. Since by Stoke's theorem the magnetic flux carried 
by the vortex is determined by the dual 
gauge field $a^{D,\tiny\mbox{tr}}_\mu$ transverse to the vortex-tangential 
and since $a^{D,\tiny\mbox{tr}}_\mu$ is -- in a covariant gauge -- 
invariant under collective boosts of the streaming 
monopoles or antimonopoles in the vortex core it 
follows that the magnetic flux solely depends 
on the monopole charge and not on the collective state of 
monopole-antimonopole motion. This, in turn, implies a center-element classification of the magnetic fluxes carried by the vortices justifying the name center-vortex loop.) Viewed on the 
level of large-holonomy calorons an unstable center-vortex loop is created within a 
region where the mean axis for the dissociation of several calorons represents a net 
direction for the monopole-antimonopole flow. Notice that each so-generated vortex 
core must form a closed loop due to the absence of isolated magnetic 
charges within the monopole condensate. In contrast to the deconfining phase a 
rotation to unitary gauge $a^{D,bg}_{\mu}=0, \varphi=|\varphi|$ is facilitated 
by a smooth, periodic gauge transformation which leaves the value ${\bf 1}$ 
of the Polyakov loop invariant: the electric $Z_2$ degeneracy, observed 
in the deconfining phase, is lifted in the ground-state. (For SU(3) it 
is an electric $Z_3$ degeneracy that is lifted.)   

\begin{figure}
\begin{center}
\leavevmode
\leavevmode
%\epsffile[80 25 534 344]{}
\vspace{5.0cm}
\includegraphics{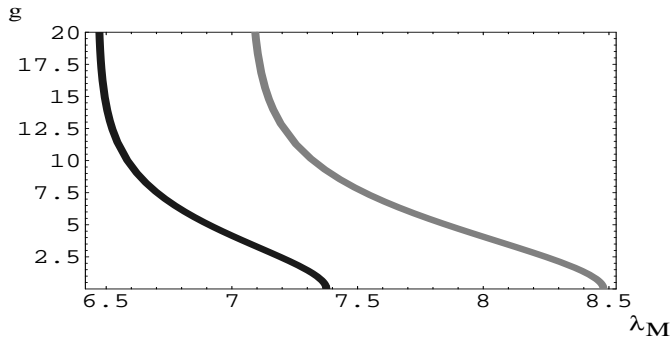}
\end{center}
\caption{The evolution of the effective magnetic gauge coupling $g$ in the preconfining 
phase for SU(2) (thick grey line) and SU(3) (thick black line) 
where $\lambda_M\equiv \frac{2\pi T}{\Lambda_M}$. At 
$\lambda_{c,M}=7.075$ (SU(2)) and $\lambda_{c,M}=6.467$ (SU(3)) $g$ diverges logarithmically, 
$g\sim -\log(\lambda_{M}-\lambda_{c,M})$. \label{g}}      
\end{figure}

By the dual (abelian) Higgs mechanism the mass of (noninteracting) 
quasiparticle modes is 
%**************
\eqb
\label{mass}
m=g|\varphi|=g|\varphi_1|=g|\varphi_2|\,.
\eqe
%**************
\noindent{\sl Evolution of magnetic coupling.} The requirement that 
Legendre transformations between thermodynamical quantities need to be invariant 
under the applied spatial coarse-graining determines the evolution of the 
magnetic coupling $g$ with temperature in terms of a first-order differential 
equation \cite{Hofmann2005}. In Fig.\,\ref{g} the temperature evolution of $g$ is shown for SU(2) and SU(3).  
%***********************
%************************
By virtue of Eq.\,(\ref{mass}) and Fig.\,\ref{g} it 
is clear that inside the preconfining phase dual gauge modes 
are massive. Thus the full Polyakov-loop expectation is extremely 
suppressed as compared to the deconfining phase: Infinitely heavy, fundamental, and fermionic 
test charges are confined by the dual Meissner effect while dual gauge modes still 
propagate with a maximal off-shellness $|\varphi|^{-1}$. 
Notice the logarithmic singularity at $T_{c,M}$. Since the energy $E_v$ and the 
pressure $P_v(r)$ of a (nonselfintersecting) center-vortex loop, whose center of mass 
is at rest with respect to the heat bath, scale like $+g^{-1}$ and $-g^{-2}$, respectively, 
this soliton starts to be generated 
as a massless and stable (spin-1/2) {\sl particle} at $T_{c,M}$ \cite{Hofmann2005}. 
Moreover, all (dual) gauge modes decouple by a diverging 
mass: For $T\le T_{c,M}$ no continuous gauge symmetry can be 
observed in the system at an extranlly applied spatial 
resolution smaller than $\varphi(T_{c,M})$.\vspace{-0.3cm}\\ 

\begin{figure}
\begin{center}
\leavevmode
\leavevmode
%\epsffile[80 25 534 344]{}
\vspace{4.5cm}
\includegraphics{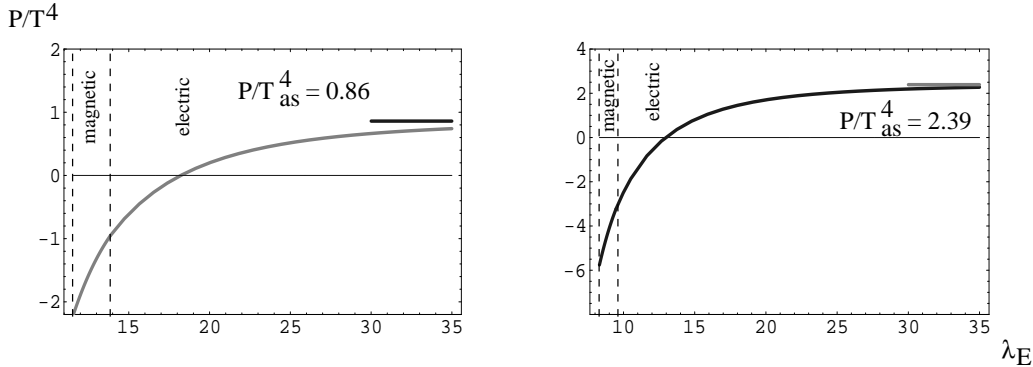}
\end{center}
\caption{\protect{$\frac{P}{T^4}$ as a function of temperature for SU(2) (left panel) and SU(3) (right panel). 
The horizontal lines indicate the respective asymptotic values, the dashed vertical lines are the phase boundaries.  
\label{pressure}}}      
\end{figure}

\begin{figure}
\begin{center}
\leavevmode
\leavevmode
%\epsffile[80 25 534 344]{}
\vspace{5.5cm}
\includegraphics{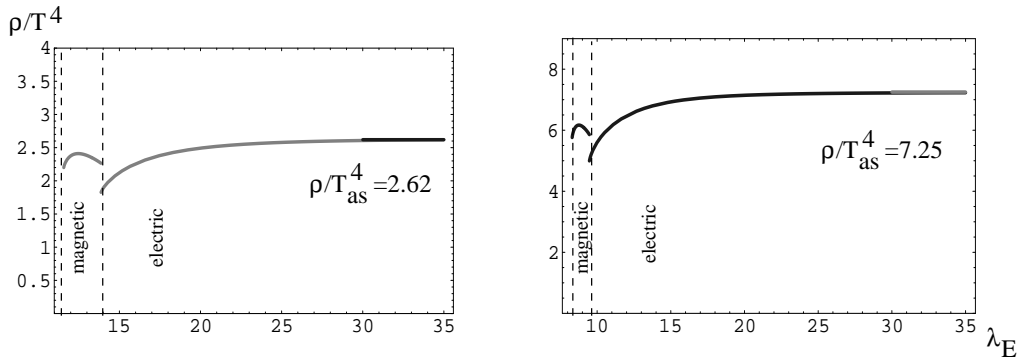}
\end{center}
\caption{$\frac{\rho}{T^4}$ as a function of temperature for SU(2) (left panel) and SU(3) (right panel). 
The horizontal lines indicate the respective asymptotic values, the dashed 
vertical lines are the phase boundaries.\label{rho}}
\end{figure}

\noindent{\sl Results and summary.}
In Figs.\,\ref{pressure} and \ref{rho} we present the results 
for the temperature evolution of the pressure and the energy 
density for both the deconfining (electric) 
and preconfining (magnetic) phase. To express the magnetic scale $\Lambda_M$ in 
terms of the electric scale $\Lambda_E$ continuity of the pressure is demanded 
across the electric-magnetic transition. We have 
$\Lambda_M\sim \left(\frac{1}{4}\right)^{-1/3}\La_E\,\ 
(\mbox{SU(2)})\,\ \mbox{and}\ \Lambda_M\sim \left(\frac{1}{2}\right)^{-1/3}\La_E\,\ 
(\mbox{SU(3)})$. 

Notice the increasing negativity of the pressure with decreasing 
temperature in the magnetic phase. On the microscopic level 
this is understood in terms of an 
increasingly large caloron-holonomy, implying an increasingly 
large repulsion of its constituent monopole and antimonopole, which, in turn, 
means an increasingly large collimation of the 
monopole-antimonopole motion in the condensate. The effect is an 
increasing rate for the creation of (unstable) 
center-vortex loops implying, after spatial 
coarse-graining, an increasingly negative ground-state 
pressure and, at the same time, an increase of the quasiparticle 
masses. At $T_{c,M}$ {\sl all} quasiparticles are infinitely 
heavy and the equation of state is $P=-\rho$, 
compare Figs.\,\ref{pressure} and \ref{rho}.  
%***********************
%************************ 
The (continuous) order 
parameter of mass-dimension one for the spontaneous breakdown of the dual gauge 
symmetry U(1)$_D$ (SU(2)) and U(1)$^2_D$ (SU(3)) across 
the (second-order like) electric-magnetic phase transition is the `photon' mass 
in Eq.\,(\ref{mass}). The associated critical exponents $\nu$ are mean-field ones, $\nu=0.5$, 
for both SU(2) and SU(3) \cite{Hofmann2005}. The transition 
is, however, not strictly second order because of 
a {\sl negative} latent heat, see Fig.\,\ref{rho}. 
The reason for this discontinuous {\sl increase} of the energy 
density across the electric-magnetic transition is the discontinuous 
increase of the number of polarizations from two to three due to the 
dual `photon' becoming massive. This effect is important because 
it stabilizes the temperature of the cosmic microwave 
background $T_{\tiny\mbox{CMB}}$, described by an SU(2) pure gauge 
theory of Yang-Mills scale $\Lambda_{\tiny\mbox{CMB}}\sim T_{\tiny\mbox{CMB}}$,   
against gravitational expansion. The latter increasingly liberates a formerly 
locked-in, spatially homogeneous Planck-scale 
axion field \cite{Hofmann2005,Wilczek2004} which, eventually, 
will drive the Universe out of thermal equilibrium globally.       

The Polyakov-loop expectation, which is an order 
parameter associated with the spontaneous breaking of the electric $Z_2$ (SU(2)) and the electric 
$Z_3$ (SU(3)) symmetry, though strongly suppressed on the magnetic side, remains 
finite across the electric-magnetic transition. This happens despite the fact that the 
magnetic ground state is nondegenerate with respect to these symmetries.     
%***********************
%************************

\section*{Acknowledgements}
The author would like to thank Mark Wise for useful conversations 
and for the warm hospitality extended to him during his 
visit to Caltech in May 2005. Financial support by 
Kavli Institute at Santa Barbara and by 
the physics department of UCLA is thankfully acknowledged.

\baselineskip25pt
\end{document}